\documentclass[twocolumn,showpacs,amsmath,amssymb,superscriptaddress,prl]{revtex4}% Physical Review Letters

\usepackage{graphicx}% Include figure files
\usepackage{dcolumn}% Align table columns on decimal point
\usepackage{bm}% bold math

\begin{document}

\title{Nonmonotonic charge occupation in double dots}

\author{J\"urgen K\"onig}
\affiliation{Institut f\"ur Theoretische Physik III, Ruhr-Universit\"at Bochum, 44780 Bochum, Germany}
\author{Yuval Gefen}
\affiliation{Department of Condensed Matter Physics, The Weizmann Institute of Science, 76100 Rehovot, Israel}

\date{\today}

\begin{abstract}
We study the occupation of two electrostatically-coupled single-level 
quantum dots with spinless electrons as a function of gate voltage.
While the total occupation of the double-dot system varies monotonically with 
gate voltage, we predict that the competition between tunneling and
Coulomb interaction can give rise to a nonmonotonic filling of the individual 
quantum dots.
This non-monotonicity is a signature of the correlated nature of the
many-body wavefunction in the reduced Hilbert space of the dots.
We identify two mechanisms for this nonmonotonic behavior, which are associated
with changes in the spectral weights and the positions, respectively, of the 
excitation spectra of the individual quantum dots. 
An experimental setup to test these predictions is proposed.
\end{abstract}
%
% 73.23.Hk Electronic transport in mesoscopic systems: 
%          Coulomb blockade; single-electron tunneling
% 73.21.La Electron states and collective excitations in multilayers, 
%          quantum wells, mesoscopic, and nanoscale systems: Quantum dots
% 73.63.Kv Electronic transport in nanoscale materials and structures: 
%          Quantum dots
% 

%\keywords{Suggested keywords}%Use showkeys class option if keyword
                              %display desired
\maketitle

{\bf Introduction.} --
Quantum dots coupled to an electron reservoir can be filled with electrons one 
by one in a controlled way.
With increasing gate voltage, the number of electrons on the dot increases in
a step-like manner.
As long as Coulomb interaction among the dot electrons is negligible, the
dot levels are filled independently from each other.
However, small semiconductor quantum dots are usually subject to strong 
Coulomb interaction: charging energy gives rise to extended plateaus of 
(almost) fixed dot charge as a function of gate voltage.
The aim of the present paper is to demonstrate that more dramatic signatures 
of Coulomb correlations can be found in studying the filling of the 
individual levels in the quantum dot. 
While the total occupation of the dot remains a monotonic function of the 
gate voltage, we predict that, under circumstances specified below, the 
individual levels are filled in a nonmonotonic way.
The physics proposed here enables one to observe non-trivial many-body 
correlations in a particularly simple (effective) 4-dimensional Hilbert space.

\begin{figure}[!ht]
\centerline{\includegraphics[width=0.6\columnwidth,angle=0]{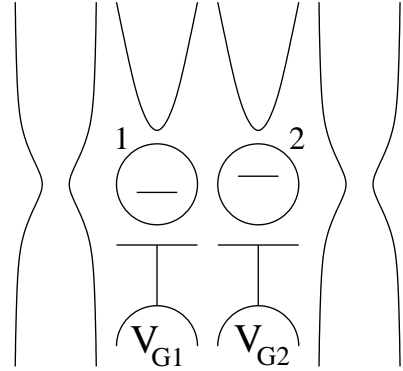}}
\caption{Two electrostatically-coupled single-level quantum dots are attached
  to electron reservoirs and gated by gate electrodes.
  The individual occupation is determined by adjacent quantum point contacts.}
\label{system_fig}
\end{figure}

To illustrate the signature and mechanisms of the nonmonotonic filling, we 
consider the simplest possible model system which allows for a separate 
access to the individual level occupation: two electrostatically-coupled 
single-level quantum dots with spinless electrons, see Fig.~\ref{system_fig}.
The double dot is equivalent to a single quantum dot accommodating two levels.
The advantage of the double-dot setup lies in the possibility to read out
separately each dot's occupation by electrostatically-coupled quantum point
contacts.
This kind of charge sensing has been experimentally demonstrated 
for single \cite{dot+qpc} and double dots \cite{2dots} recently.
The quantum dots are tunnel coupled to one common or to two separate electron 
reservoirs.

We model the double-dot system with the standard tunnel Hamiltonian
$H = H_{\rm dot 1}+H_{\rm dot 2}+H_{\rm ch}+H_{\rm lead 1}+H_{\rm lead 2}
+H_{\rm T,1}+H_{\rm T,2}$.
Quantum dot $i=1,2$, described by $H_{{\rm dot}, i}=\epsilon_i c^\dagger_ic_i$,
accommodates a level with energy $\epsilon_i$, measured relative to the Fermi 
energy in the leads.
The level energies can be tuned by the gate electrodes.
Without loss of generality, we always assume $\epsilon_2 \ge \epsilon_1$.
The charging energy is accounted for by $H_{\rm ch}= U n_1 n_2$.
Each dot is coupled to an electron reservoir \cite{com1}, 
$H_{{\rm lead},i}=\sum_k \epsilon_{ki} a^\dagger_{ki} a_{ki}$.
Tunneling is modeled by $H_{{\rm T}, i}= \sum_{k} \left( t_{i} 
c^\dagger_i a_{ki} + h.c. \right)$, where we assume the tunnel matrix elements 
to be energy independent.
The tunnel coupling introduces a line width $\Gamma_i = 2\pi |t_i|^2 \rho_i$, 
where $\rho_i$ is the density of states of lead $i$ at the Fermi energy.

There are four possible states for the double-dot system: the double dot
being empty ($\chi=0$), singly ($\chi=1,2$), or doubly ($\chi=d$) occupied.
The corresponding energies are $E_\chi$ \cite{comx}.
To address the question of how the occupations $\langle n_1\rangle$ and
$\langle n_2\rangle$ of the dots vary as both levels $\epsilon_1$ and 
$\epsilon_2$ are simultaneously pulled down, we keep the bare level separation 
$\Delta=\epsilon_2 - \epsilon_1$ fixed.
The fact that tunneling does not commute with the kinetic and charging terms 
is essential for the physics outlined below.

{\bf Regimes.} --
We begin with identifying the regime at which a nonmonotonic filling of the 
individual dots are expected.
Consider first $U=0$.
The spectral density of dot $i$ is a Lorentzian centered around $\epsilon_i$ 
with a width $\Gamma_i$.
As $\epsilon_i$ crosses the Fermi energy of the lead, the occupation
$\langle n_i \rangle$ of level $i$ changes from $0$ to $1$.
The width of this transition is governed by $\max \{ \Gamma_i, k_{\rm B}T \}$.
It is obvious that in this case the occupation of each level is a monotonic 
function of the gate voltage.
We, thus, turn our attention to the limit 
$U \gg \max\{ \Delta,\Gamma,k_{\rm B}T\}$.

For temperatures larger than the level broadening, $k_{\rm B}T \gg \Gamma$, 
the influence of the tunnel couplings on the occupations is negligible and
the double-dot system is filled according to the Boltzmann factors 
$\exp(-E_\chi^{(0)}/k_{\rm B}T)$ of the corresponding bare energies 
$E_\chi^{(0)}$ with $\epsilon_i = E_i^{(0)} - E_0^{(0)}$ and 
$U= E_{\rm d}^{(0)} - E_1^{(0)} - E_2^{(0)} + E_0^{(0)}$.
Now, the occupations of the two dots are correlated due to charging energy,
$\langle n_1 n_2 \rangle - \langle n_1 \rangle \langle n_2 \rangle < 0$, but 
the individual dot occupations $\langle n_i\rangle$ are still found to depend 
monotonically on the gate voltage.
Therefore, we require $\Gamma \gtrsim k_{\rm B}T$.

If the level separation is small as compared to either level width or 
temperature, $\max \{ k_{\rm B} T , \Gamma \} \gg \Delta$, then
the double dot will be filled in a symmetric way,
$\langle n_1\rangle \approx \langle n_2\rangle$, which again yields a 
monotonic filling.
This restricts the regime of interest to 
$\Delta \gtrsim \Gamma \gtrsim k_{\rm B}T$.

\begin{figure}[!ht]
\centerline{\includegraphics[width=0.8\columnwidth,angle=0]{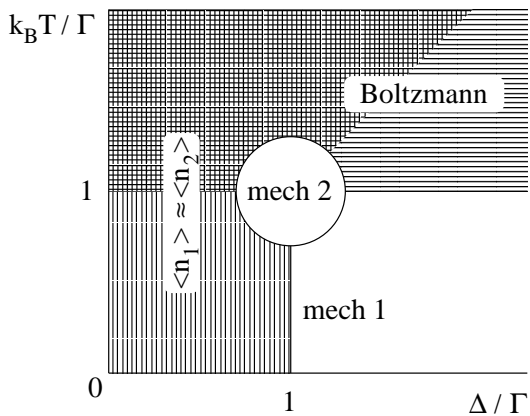}}
\caption{For $k_{\rm B}T \gg \Gamma$ or $\Gamma \gg \Delta$, the occupation
  of the individual dots is a monotonic function of the gate voltage.
  Nonmonotonicities due to mechanism 1 are strongest for 
  $\Delta \sim \Gamma \gg k_{\rm B}T$, and mechanism 2 becomes effective for
  $\Delta \sim \Gamma \sim k_{\rm B}T$ and $\Gamma_1 \neq \Gamma_2$.}
\label{regimes_fig}
\end{figure}

{\bf Mechanisms for nonmonotonicity.} --
We identify two different mechanisms by which a nonmonotonic filling of the 
individual levels can occur.
They are both related to the Coulomb interaction, with two different 
consequences concerning the spectral density.
First, as mentioned above, there is spectral weight of level $2$ not only 
around $\epsilon_2$ (describing filling or depleting of dot $2$ while dot $1$
is empty) but also around $\epsilon_2+U$ (describing transitions when dot $1$
is filled).
The total spectral weight is normalized to $1$, with the relative spectral 
weight between the two peaks being determined by occupation of dot $1$.
Both peaks of level $2$ are broadened by $\Gamma_2$ \cite{com2}, which yields 
a partial filling of the dot once the peak is close to the Fermi energy.
As the level in dot $1$ passes through the Fermi level and dot $1$ is filled,
spectral weight for the level in dot $2$ is transfered from the peak 
around $\epsilon_2$ to the one around $\epsilon_2+U$, which is further away 
from the Fermi level, yielding a reduction of the partial occupation of 
dot $2$, giving rise to a nonmonotonic filling.
This mechanism will be visible for $\Delta \gg k_{\rm B}T$ (since otherwise 
the filling is dominated by thermal fluctuations) and is most pronounced
for $\Delta \sim \Gamma \gg k_{\rm B}T$.

The second mechanism is based on a renormalization of the peak positions in 
the spectral density due to Coulomb interaction \cite{haldane}.
As discussed in more detail below, tunneling in and out of dot $2$ yields
a renormalization of dot level $1$, $\epsilon_1 \rightarrow \tilde \epsilon_1$.
The renormalization is strongest when level $2$ is close to the Fermi energy.
This renormalization can give rise to a nonmonotonic filling of dot $2$
under the condition $\Delta \sim k_{\rm B}T$ and 
$\langle n_1 \rangle + \langle n_2 \rangle \approx 1$.
In this case, both dots will be partially filled once $\tilde \epsilon_1$ and
$\tilde \epsilon_2$ are below the Fermi level.
The relative occupation is approximately given by $\exp(-\Delta/k_{\rm B}T)$.
A renormalization of the level splitting $\Delta$ then changes the relative 
occupation.
If level $2$ is close to the Fermi level, then level $1$ is strongly 
renormalized towards higher energies.
As a consequence, the occupation of dot $2$ is increased on the cost of
level $1$, i.e., there is a reshuffling of the relative occupation. 
When the gate voltage is increased further, the renormalization becomes 
weaker, and level 1 gains back some of the occupation it lost to level 2, 
which leads to a nonmonotonicity of $\langle n_2 \rangle$.
It turns out that for symmetric coupling, $\Gamma_1=\Gamma_2=\Gamma$, the 
renormalization of the level separation is negligible in the regime 
$\Delta \sim \Gamma \sim k_{\rm B}T$ since both levels are renormalized 
approximately by the same amount.
To get a nonmonotonic dot filling, an asymmetric coupling 
$\Gamma_1 \neq \Gamma_2$ is required.

{\bf Perturbation expansion.} --
To substantiate the qualitative ideas outlined above, we perform a
perturbation expansion of the dot occupations in the tunnel-coupling strength
$\Gamma$.
For this, we employ a diagrammatic imaginary-time technique that was introduced
to study charge fluctuations in a metallic single-electron box \cite{grabert},
adjusted to a two single-level-dot system.
The central quantity is the partition function 
$Z = \sum_{\chi} \exp (-\beta E_\chi)$.
The corresponding effective energies 
$E_\chi = E_\chi^{(0)} + E_\chi^{(1)} + {\cal O}(\Gamma^2)$
can be expanded in orders of $\Gamma$.
Evaluation of $E_\chi^{(1)}$ yields
$E_0^{(1)} = - [ \sigma_1 (\epsilon_1) + \sigma_2 (\epsilon_2) ]$,
$E_1^{(1)} = - [ \sigma_1 (\epsilon_1) + \sigma_2 (\epsilon_2+U) ]$,
$E_2^{(1)} = - [ \sigma_1 (\epsilon_1+U) + \sigma_2 (\epsilon_2) ]$, and
$E_{\rm d}^{(1)} = - [ \sigma_1 (\epsilon_1+U) + \sigma_2 (\epsilon_2+U) ]$,
with 
$\sigma_i(\omega) = \frac{\Gamma_i}{2\pi} \left[ \ln \left( 
\frac{\beta D}{2\pi} \right) - {\rm Re} \, \Psi \left( 
\frac{1}{2} + i \frac{\beta\omega}{2\pi} \right) \right]$,
where the bandwidth $D$ appears as a high-energy cutoff at this intermediate 
step but drops out for all physical observables since they only depend on
differences of $E_\chi$'s.
As a result, the dot levels are renormalized according to
$\tilde \epsilon_i = \epsilon_i + \epsilon_i^{(1)}$ with
\begin{equation}
\label{eps_1}
  \epsilon_1^{(1)} = \frac{\Gamma_2}{2\pi} {\rm Re}
  \left[ \Psi \left( \frac{1}{2} + i \frac{\beta(\epsilon_2+U)}{2\pi} \right)
    - \Psi \left( \frac{1}{2} + i \frac{\beta \epsilon_2}{2\pi} \right)
    \right] \, , 
\end{equation}
for $\epsilon_2^{(1)}$ the same expression holds with $1$ replaced by $2$, 
and $U$ remains unrenormalized in first order in $\Gamma$.
Here, $\Psi(x)$ is the digamma function.
From Eq.~(\ref{eps_1}) we see that the renormalization 
of the level in dot $1$ is proportional to the tunnel coupling of dot $2$
(and vice versa).
It is maximal when $\epsilon_2$ or $\epsilon_2+U$ is in resonance with the 
Fermi level, and vanishes in the absence of Coulomb charging.

Thermodynamic quantities such as the occupation of the dots are obtained by
adequate logarithmic derivatives of the partition function.
In particular, 
\begin{eqnarray}
\label{ni}
  \langle n_i \rangle &=& - \frac{1}{\beta} \frac{\partial}
	  {\partial \epsilon_i} \ln Z
\\
\label{n1n2}
  \langle n_1 n_2 \rangle &=& - \frac{1}{\beta} \frac{\partial}
	  {\partial U} \ln Z \, .
\end{eqnarray}
The first-order corrections of these quantities are directly obtained from
\begin{eqnarray}
\label{z}
  \frac{\left( \ln Z \right)}{k_{\rm B}T}^{(1)} \!\! = \left\{
    \left[1-\langle n_2 \rangle ^{(0)}\right] \sigma_1(\epsilon_1) 
    + \langle n_2 \rangle ^{(0)} \sigma_1(\epsilon_1+U)
    \right. \nonumber \\  \left. 
    + \left[ 1-\langle n_1 \rangle ^{(0)} \right] \sigma_2(\epsilon_2) 
    + \langle n_1 \rangle^{(0)} \sigma_2(\epsilon_2+U)
    \right\} \,
\end{eqnarray}
plus a constant that is independent of $\epsilon_1$, $\epsilon_2$, and $U$.
The set of Eqs.~(\ref{ni})-(\ref{z}) provides the starting point for the
subsequent quantitative analysis.
For example, the correction to occupation of dot 2 is 
\begin{eqnarray}
\label{n2}
  \langle n_2 \rangle^{(1)} &=& - \left[ 1 - \langle n_1 \rangle ^{(0)}\right] 
  \frac{\partial \sigma_2(\epsilon_2)}{\partial \epsilon_2} 
  - \langle n_1 \rangle ^{(0)} 
  \frac{\partial \sigma_2(\epsilon_2 +U)}{\partial \epsilon_2} 
  \nonumber \\
  && + \frac{\partial \langle n_2 \rangle ^{(0)}}{\partial \epsilon_1}
  \left[ \sigma_2(\epsilon_2) - \sigma_2(\epsilon_2+U) \right]
  \nonumber \\
  && + \frac{\partial \langle n_2 \rangle ^{(0)}}{\partial \epsilon_2}
  \left[ \sigma_1(\epsilon_1) - \sigma_1(\epsilon_1+U) \right] \, ,
\end{eqnarray}
where we have used 
$\partial \langle n_2 \rangle ^{(0)} / \partial \epsilon_1 =
\partial \langle n_1 \rangle ^{(0)} / \partial \epsilon_2$.

\begin{figure}[!ht]
\vspace*{.14cm}
  \centerline{\includegraphics[width=0.8\columnwidth,angle=0]{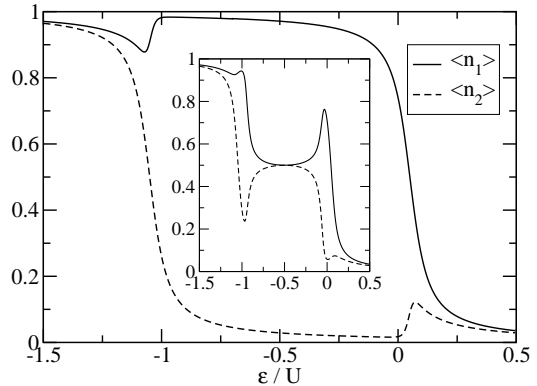}}
  \caption{Nonmonotonic dot filling due to mechanism 1. 
  The parameters are $\Delta = 0.1U$, $\Gamma_1 = \Gamma_2 = 0.1 U$, 
  and $k_{\rm B}T = 0.01 U$ \cite{com_fig}.
  Inset: Result for the symmetrized Hartree approximation.}
  \label{mech_1}
\end{figure}

{\bf Results.} --
In Fig.~\ref{mech_1} we show the dot occupations as a function of
the mean level position $\epsilon = (\epsilon_1+\epsilon_2)/2$ in the regime
$\Delta \sim \Gamma \gg k_{\rm B}T$.
We find nonmonotonicities which we relate to mechanism 1 discussed above.
For an approximate analytical understanding of the result, we concentrate on
the nonmonotonicity of $\langle n_2 \rangle$ on the right step.
Since $\epsilon_2 > \epsilon_1$ and $\Delta \gg k_{\rm B}T$, we can set 
$\langle n_2 \rangle ^{(0)} \approx 0$ for the whole region over which
level $1$ is filled up, and $\langle n_1 \rangle ^{(0)}$ is independent of
$\epsilon_2$.
Nevertheless, level $2$ is partially filled due quantum fluctuations,
\begin{equation}
  \langle n_2 \rangle^{(1)} \approx
    \left[ 1-\langle n_1 \rangle^{(0)} \right] 
    \frac{\Gamma_2}{2\pi \epsilon_2} + 
    \langle n_1 \rangle^{(0)} \frac{\Gamma_2}{2\pi (\epsilon_2 + U)} \, .
\end{equation}
As level $1$ passes through the Fermi level, filling up quantum dot $1$, 
the occupation of dot $2$ drops from $\Gamma_2/ (2\pi \epsilon_2)$ down to 
$\Gamma_2/[ 2\pi (\epsilon_2+U)]$.
Within our perturbative analysis, the width of this drop is provided by 
temperature, as a broadening of level 1 due to quantum fluctuations would 
enter in higher orders only.
When going to higher order in $\Gamma$, the width is expected to be
$\max \{ k_{\rm B}T, \Gamma_1 \}$.
The amplitude of the drop is approximately
\begin{equation}
  \delta \langle n_2 \rangle \approx \frac{\Gamma_2 U}{2\pi \Delta (\Delta +U) } \, .
\end{equation}

\begin{figure}[!ht]
\vspace*{.48cm}
  \centerline{\includegraphics[width=0.8\columnwidth,angle=0]{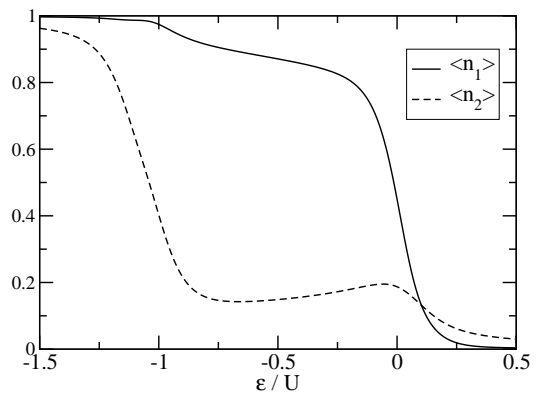}}
  \caption{Nonmonotonic dot filling due to mechanism 2 for $\Delta = 0.1U$, 
    $\Gamma_1 = 0.01 U$, $\Gamma_2 = 0.1 U$, and $k_{\rm B}T = 0.05 U$.}
  \label{mech_2}
\end{figure}

We now turn to the regime $\Delta \sim \Gamma \sim k_{\rm B}T$ to discuss
mechanism 2.
From inspection of Eq.~(\ref{eps_1}) we realize that in the given regime
the renormalization of both levels is roughly the same.
In order to generate a substantial renormalization of the level separation
that can give rise to a nonmonotonicity, we choose different coupling strengths
$\Gamma_1 \neq \Gamma_2$.
In the following we discuss the case $\Gamma_2 > \Gamma_1$.
The result is shown in Fig.~\ref{mech_2}.
The second line of Eq.~(\ref{n2}) is associated with the renormalization of 
level $1$.
The renormalization (towards higher energies) is strongest when $\epsilon_2$
passes through the Fermi energy.
The peak of the nonmonotonicity is, therefore, located to the left as compared 
to the one related with mechanism 1. 
The width is given by $\Gamma_2$.
For an estimate of its height $\delta \langle n_2 \rangle$, we neglect 
both the possibility of the double dot being empty or doubly occupied, i.e., 
$\langle n_2 \rangle ^{(0)} \approx \exp(-\beta \Delta) / 
[\exp(-\beta \Delta)+1]$, and we assume $U \gg k_{\rm B}T$.
We find
\begin{equation}
  \delta \langle n_2 \rangle \approx
  \frac{\beta e^{-\beta\Delta}}{(e^{-\beta\Delta}+1)^2} 
  \frac{\Gamma_2 - \Gamma_1}{2\pi} \ln \frac{\beta U}{2\pi}
  \, .
\end{equation}

If we reverse the asymmetry of the coupling strengths, $\Gamma_1 > \Gamma_2$,
we find, by using analogous arguments as above, a nonmonotonicity of
$\langle n_1\rangle$ on the left step, while $\langle n_2\rangle$ remains
monotonic.

\begin{figure}[!ht]
\vspace*{.4cm}
\centerline{\includegraphics[width=0.8\columnwidth,angle=0]{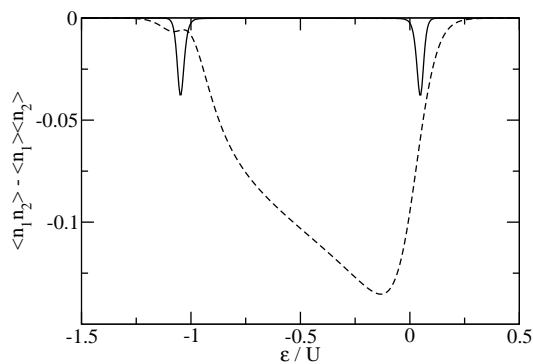}}
  \caption{Correlator $\langle n_1 n_2 \rangle - \langle n_1 \rangle 
    \langle n_2 \rangle$.
    The parameters are as in Fig.~\ref{mech_1} (solid line) and
    Fig.~\ref{mech_2} (dashed line).}
  \label{corr}
\end{figure}

In Fig.~\ref{corr} we show the correlator 
$\langle n_1 n_2 \rangle - \langle n_1 \rangle \langle n_2 \rangle$, expanded
up to first order in $\Gamma$, for the two set of parameters chosen in
Fig.~\ref{mech_1} and Fig.~\ref{mech_2}.
We find that the nonmonotonicity of the dot occupation is accompanied with
an enhancement of correlation.

Throughout our analysis, we assumed flat bands, i.e., 
particle-hole-symmetric densities of states, in the leads.
As a consequence, $\langle n_1 \rangle$ at $\epsilon$ for given 
$\Gamma_1, \Gamma_2$ is identical to $1- \langle n_2 \rangle$ at
$U-\epsilon$ with the values of $\Gamma_1$ and $\Gamma_2$ being interchanged.
Thus, our discussion, which was focused on the right step of 
$\langle n_2 \rangle$ holds for the left step of $\langle n_1 \rangle$ as well
(after interchanging $\Gamma_1 \leftrightarrow \Gamma_2$).

In parallel to this perturbation theory, we have performed an 
equation-of-motion (EOM) analysis up to the second hierarchy
solving a set of 11 coupled equations for the various Green's functions.
This approach, however, is not systematically controllable.
Truncating the EOMs on the first hierarchy results in the symmetrized Hartree
approximation \cite{hartree} with the Green's function 
$G_{11}^{\rm ret} (\omega)$ given by
$[G_{11}^{\rm ret} (\omega)]^{-1} = [ \left( 1- \langle n_2 \rangle \right) / 
(\omega - \epsilon_1) + \langle n_2 \rangle / (\omega - \epsilon_1 - U)]^{-1} 
+ i\Gamma_1/2$ and similar for $G_{22}^{\rm ret} (\omega)$.
The resulting $\langle n_1 \rangle$ and $\langle n_2 \rangle$ tend to be biased
towards each other, but the nonmonotonicity is still observed, see the
inset of Fig.~\ref{mech_1}.

{\bf Summary.} --
In summary, we predict that the competition between tunneling and charging
energy can give rise to nonmonotonic filling of individual quantum-dot levels.
We identify two different mechanisms leading to nonmonotonicities and determine
the regimes at which they occur.
Based on a perturbation expansion in the tunnel couplings, we derive analytic 
expressions for the location and the strength of the signal.

{\bf Acknowledgements.} --
We thank Y. Alhassid, B. Altshuler, R. Berkovits, M. Kiselev, J. Martinek, 
Y. Oreg, A. Silva, M. Sindel, and J. von Delft for useful discussions.
This work was supported by the DFG through SFB 491 and GRK 726, the US-Israel 
BSF, the ISF of the Israel Academy of Science, the Alexander von Humboldt 
foundation (through the Max-Planck Award) and the EC HPRN-CT-2002-00302-RTN.
While performing this work, we became aware of parallel work \cite{sindel} in
which nonmonotonic level occupations are discussed as well.


\begin{thebibliography}{99}
\bibitem{comx}
Here, $E_\chi$ is defined as the expectation value of the Hamiltonian in 
the subspace $\chi$.
\bibitem{com_fig}
The perturbative result for $\langle n_1 \rangle$ breaks down at low 
temperature when $\epsilon_2 \approx 0$.
Therefore, in the plot we replace $\langle n_1 \rangle^{(0)} + 
\langle n_1 \rangle^{(1)}$ in this region by the occupation of a noninteracting
level.
This does not affect $\langle n_2 \rangle$, for which the perturbation scheme
remains valid.
The same procedure is done for $\langle n_2 \rangle$ around 
$\epsilon_1 \approx 0$.
\bibitem{hartree}
This is to be distinguished from the self-consistent Hartree approximation, in 
which the bare level energies $\epsilon_1$ and $\epsilon_2$ are replaced by 
$\epsilon_1 + U \langle n_2 \rangle$ and $\epsilon_2 + U \langle n_1 \rangle$.
Within that scheme, the spectral density for each dot has {\it one} peak that 
is renormalized {\it linearly} with $U$.
In contrast, our understanding of the nonmonotonic filling at low temperature 
(mechanism 1) is based on the fact that the spectral density of each dot has 
{\it two} peaks.
Mechanism 2, on the other hand, which is effective at higher temperatures, 
is related to a renormalization of the level positions that depend
{\it logarithmically} on $U$.
\end{thebibliography}
\end{document}